\documentclass[preprint,prd,showpacs,preprintnumbers,amsmath,amssymb,floatfix]{revtex4}

\usepackage{amsmath}
\usepackage{graphicx}
\usepackage{dcolumn}
\usepackage{bm}
\usepackage{ulem} 
\usepackage[usenames]{color}
\usepackage{epstopdf}
\usepackage{epsfig}
\usepackage{float}
\usepackage{subfigure}
\usepackage{multirow}
\usepackage[colorlinks,linkcolor=blue,anchorcolor=teal,citecolor=red]{hyperref}

\usepackage{array}

\newcommand{\nc}{\newcommand}       
\nc{\vc}[1] {\mbox{\boldmath $#1$}} 
\nc{\del}       {\partial}              
\nc{\bra}       {\langle}               
\nc{\ket}       {\rangle}               
\nc{\bras}[1]   {\langle #1|}           
\nc{\kets}[1]   {|#1\rangle}            
\nc{\mapleft}[1]{           
	\smash{\mathop{\,          %
			\hbox to 1.5cm{\rightarrowfill}\, }\limits_{#1}}}
\nc{\nn}      {\\\nonumber} \nc{\vs}      {\vspace{-0.275cm}}
\nc{\fra}    {\frac{1}{2}}
\nc{\mb}        {\mathbf}

\begin{document}
	
	\preprint{}
	\title{The first-order phase transition in the neutron star from the deep neural network} 
	
	\author{Wenjie Zhou}
	\affiliation{School of Physics, Nankai University, Tianjin 300071,  China}
	\author{Jinniu Hu}
	\email{hujinniu@nankai.edu.cn}
	\affiliation{School of Physics, Nankai University, Tianjin 300071,  China}
	\affiliation{Shenzhen Research Institute of Nankai University, Shenzhen 518083, China}
	\author{Ying Zhang}
	\affiliation{Department of Physics, Faculty of Science, Tianjin University,Tianjin 300072, China}
	\author{Hong Shen}
	\affiliation{School of Physics, Nankai University, Tianjin 300071,  China}
	\date{\today} 
   \begin{abstract}
    This study investigates the first-order phase transition within neutron stars, leveraging the deep neural network (DNN) framework alongside contemporary astronomical measurements. The equation of state (EOS) for neutron stars is delineated in a piecewise polytropic form, with the speed of sound ($c_s$) serving as a pivotal determinant. In the phase transition region, $c_s$ is presumed to be zero, while in other intervals, it is optimized utilizing the DNN. Various onset energy densities of phase transition ($\varepsilon_{pt}$), spanning from $2\varepsilon_0$ to $3\varepsilon_0$ (where $\varepsilon_0$ denotes the energy density at nuclear saturation density), as well as phase transition widths ($\Delta\varepsilon$) ranging from $0.5\varepsilon_0$ to $\varepsilon_0$, are examined.  Our findings underscore that smaller values of $\varepsilon_{pt}$ lead to a more substantial impact of $\Delta\varepsilon$ on neutron star properties, encompassing maximum mass, corresponding radius, tidal deformability, phase transition mass, and trace anomaly. Conversely, when $\varepsilon_{pt}$ exceeds $2.5\varepsilon_0$, the influence of $\Delta\varepsilon$ diminishes, resulting in a stiffer EOS compared to scenarios lacking a phase transition. Furthermore, the trace anomaly at high density shifts to negative values upon the commencement of the phase transition. It is noteworthy that the correlations between the average speed of sound at different energy density segments demonstrate a notably weak connection. The discernment of whether a phase transition has occurred with the present observables of neutron stars poses a challenging task.
   \end{abstract}
	
	\keywords{Neutron star, Machine learning, Gaussian process regression}
	
	\maketitle
\section{Introduction}\label{sec1}
The neutron star, arising as a consequence of a supernova explosion from a massive star at the end of its life, can be considered as a natural laboratory for exploring the phase diagram predicted by quantum chromodynamics (QCD) theory \cite{lattimer2007,oertel2017}. Inside the core of neutron stars, the density of strong interaction matter can reach approximately $5$ to $10$ times nuclear saturation density ($\rho_0$), while the temperature is presumed to approach zero as the star cools down \cite{fukushima2010}. Consequently, exotic baryons such as $\Lambda$, $\Sigma$, and $\Xi$ hyperons, along with deconfined quarks, may exist within the neutron star's core based on our current understanding \cite{xing2017,hu2017,zhang2018,huang2022}.

In the past decade, significant advancements have been made in astronomical measurements, particularly concerning pulsars, known as rotating neutron stars emitting precise pulses of radiation. Importantly, several massive neutron stars were detected using the Shapiro time delay effect, including PSR J1614-2230 ($M=1.908\pm0.016 M_\odot$) \cite{demorest2010,fonseca2016,arzoumanian2018}, PSR J0348+0432 ($M=2.01\pm0.04 M_\odot$) \cite{antoniadis2013}, and PSR J0740+6620 ($M=2.08\pm0.07 M_\odot$) \cite{cromartie2020}. The masses and radii of two neutron stars (PSR J0030+0451 and PSR J0740+6620) were simultaneously measured by the Neutron star Interior Composition Explorer (NICER), installed on the international space station, enabling the detection of soft X-rays emitted from hotspots on the neutron star's surface \cite{riley2019,miller2019,riley2021,miller2021}. In 2022, PSR J0952-0607, the fastest rotating and heaviest neutron star, was reported, boasting a mass of $M=2.35\pm0.17 M_\odot$ and a rotation frequency of $709$ Hz \cite{romani2022}. Additionally, the possibility of neutron stars existing as the primary compact objects in the GW190814 event, with a mass of $2.59^{+0.08}_{-0.09} M_\odot$ \cite{abbott2020gw190814}, and as companion in the PSR J0514-4002E, with a mass of $2.31^{+0.41}_{-0.22} M_\odot$ \cite{barr2024}, cannot be disregarded.

Moreover, the detection of gravitational waves from the first binary neutron star merger, known as the GW170817 event \cite{abbott2017gw170817}, has illuminated the neutron star's tidal deformability—another crucial property, besides mass and radius for constraining the equation of state (EOS) and related parameters such as the second Love number. The dimensionless tidal deformability of a neutron star at $1.4 M_\odot$ was inferred as $\Lambda_{1.4}=190^{+390}_{-120}$ from the GW170817 event \cite{abbott2018macfoy}.

The mass, radius, and tidal deformability of neutron stars are primarily governed by the EOS of strong interaction matter, reflecting the relationship between pressure and energy density \cite{lattimer2000,weber2005,baym2018}, $P(\varepsilon)$, through the Tolman–Oppenheimer–Volkoff (TOV) equation \cite{tolman1939,oppenheimer1939}. Traditionally, the EOS of neutron stars is segmented into various regions, including the outer crust, inner crust, outer core, and inner core. While the EOSs for the outer crust, inner crust, and outer core are derived from nucleon and lepton interactions, resulting in finite nuclei and infinite nuclear matter, the inner core of neutron stars, characterized by extremely high densities, may including hyperons and quarks \cite{lattimer2001}. Significantly, the discourse on the hadron-quark phase transition in the inner core of massive neutron stars have piqued substantial interest \cite{glendenning2001,baym2018,chatziioannou2020,somasundaram2023}.

Due to the complexities of QCD theory, it remains unclear how the hadron phase transitions to the quark phase within neutron stars. Three commonly employed schemes were utilized to investigate potential hadron-quark transformations. The first scheme treats the transition as a first-order phase transition, often described using constructions such as the Gibbs or Maxwell construction, or the hadron-quark pasta phase \cite{klahn2007,agrawal2010,orsaria2013,wu2017,wu2018,ju2021a,ju2021b}. The second scheme considers the mixing region between the hadron and quark phases as a hadron-quark crossover, where the EOS is obtained by interpolating between the EOSs of pure hadron and quark matter \cite{masuda2013a,masuda2013b,kojo2015,kojo2022,huang2022}. Recently, McLerran and Reddy introduced the concept of quarkyonic matter as the third scheme, where a Fermi sphere of quarks is enveloped by a nucleon shell in momentum space \cite{mclerran2019,han2019,zhao2020,fujimoto2024}. Notably, the first-order phase transition between hadron and quark phases does not stiffen the EOS, whereas the latter two schemes hold potential for forming more massive neutron stars.

With the proliferation of neutron star observables, data-driven methodologies have become prevalent in EOS investigations. These methodologies, which often involve the use of Bayesian inference or deep neural networks (DNNs), aim to infer or train the EOS based on parameterized or non-parametric formulations \cite{ozel10, raithel17, miao21, fujimoto18,fujimoto20,fujimoto21, farrell22,ferreira22,landry19,essick20a,essick20b,ferreira21,murarka22,zhou2024}. However, uncertainties continue to linger in determining the phase-transition density and intervals, given the limited sensitivity of neutron star mass-radius observations to these quantities \cite{takatsy2023,fujimoto2023,gorda2023}.

Fujimoto et al. spearheaded a machine learning approach utilizing DNNs to generate the EOS of the neutron star, based on training data from joint probability distributions of mass and radius observables \cite{fujimoto18,fujimoto20,fujimoto21}. The EOS is parameterized in a piecewise polytropic form, incorporating a set of speed of sound parameters, $\langle c_{s,i}^2\rangle$, as the output layer of the DNN. We extend this approach with a non-parametric EOS generated through Gaussian processes \cite{zhou2023}. Interestingly, during the DNN generation process, we observed instances where the speed of sound at certain density segments tended towards zero, corresponding to the Maxwell construction of first-order phase transitions—an aspect not previously included in training data.

Consequently, we utilize DNNs to generate EOS with first-order phase transitions, under constraints from the most recent neutron star observations, while manipulating the onset density and width of phase transitions within fixed boundaries. We explore their impacts on the mass-radius relation and tidal deformability of neutron stars. The structure of this paper is as follows: Section \ref{sec2} presents the methodology of DNNs for EOS, Section \ref{sec3} provides numerical results and discussions, and Section \ref{sec4} offers a summary and perspectives.

\section{Deep neural network methodology for neutron star}\label{sec2}

In this study, we employed utilized the DNN framework proposed by Fujimoto et al.~\cite{fujimoto20,fujimoto21} to investigate phase transitions in neutron stars. The dense matter's EOS is depicted via a piecewise-polytropic formulation \cite{read2009},
\begin{equation}
	p(\varepsilon)=K_i\varepsilon^{\gamma_i}, ~~~\varepsilon_{i-1}< \varepsilon < \varepsilon_i.
	\end{equation}
 The pressure $p_i$ at the endpoints of each energy density segment is computed by $p_i = p_{i-1} + \langle c_{s,i}^2\rangle (\varepsilon_i-\varepsilon_{i-1})$, where $\langle c_{s,i}^2\rangle$ denotes the average speed of sound of the dense matter in the $i$-th segment,
\begin{equation}
	\langle c_{s,i}^2\rangle=\int^{\varepsilon_i}_{\varepsilon_{i-1}}\frac{d\varepsilon}{\varepsilon_i-\varepsilon_{i-1}}c^2_s=\frac{1}{\varepsilon_i-\varepsilon_{i-1}}\int^{p_i}_{p_{i-1}}=c^2_{s,i}.
	\end{equation}

 The energy density is divided into five intervals: [$\varepsilon_0$, 2$\varepsilon_0$], [2$\varepsilon_0$, 4$\varepsilon_0$], [4$\varepsilon_0$, 6$\varepsilon_0$], [6$\varepsilon_0$, 8$\varepsilon_0$], and [8$\varepsilon_0$, 10$\varepsilon_0$], with $\varepsilon_0$ signifying the energy density at nuclear saturation density. The EOS below $\varepsilon_0$ is adopted the one from the SLy4 set including the crust region \cite{sly4}.
 
Based on prior studies, the phase transition in a neutron star is postulated to take place within the interval [2$\varepsilon_0$, 4$\varepsilon_0$]. In this study, we focus on discussing the first-order phase transition with Maxwell construction, wherein the pressure remains constant and the speed of sound within the phase transition region is set to zero. To delineate this transition, two parameters, $\varepsilon_{pt}$ and $\Delta\varepsilon$, are manually chosen. The implementation of this approach is designed to assuage the potential adverse ramifications of excessive variability on the convergence of the DNN model.

The interval [2$\varepsilon_0$, 3$\varepsilon_0$] is evenly subdivided into ten smaller intervals, culminating in eleven terminal points that symbolize potential phase transition commencement points. The length of the phase transition segment is considered to be either $0.5\varepsilon_0$ or $1.0\varepsilon_0$. Consequently, within the interval [2$\varepsilon_0$, 4$\varepsilon_0$], besides the phase transition segment, one or two segments devoid of phase transition may be present. In instances where the phase transition initiates at $2\varepsilon_0$, no preceding interval exists; conversely, if the phase transition culminates at $4\varepsilon_0$, no following interval exists.

Ultimately, the energy density range [$\varepsilon_0$, $10\varepsilon_0$] is divided into either $6$ or $7$ intervals. Excluding the phase transition segment, the EOS comprises $5$ or $6$ segments predominantly influenced by the average speed of sound $\langle c_{s,i}^2\rangle$.

The generation of the EOS using the DNN relies on the segmentation of the energy density. Within each segment, the average speed of sound $\langle c_{s,i}^2\rangle$ is randomly {within the uniform distribution} chosen from the interval $[0.01, 0.99]$ (with the speed of light $c=1$). Once $\langle c_{s,i}^2\rangle$ is determined, the exponent and coefficient of the polytropic function for each segment, denoted as $\gamma_i$ and $K_i$ respectively, are calculated using the equations:
\begin{equation}
	\gamma_i=\frac{\ln(p_i/p_{i-1})}{\ln(\varepsilon_i/\varepsilon_{i-1})}, ~~~~K_i=\frac{p_i}{\varepsilon^{\gamma_i}_i}.
	\end{equation}

Thus, the resultant EOS of neutron stars is determined by the average speed of sound. It is important to note that while the average speed of sound, $\langle c_{s,i}^2\rangle$ may be less than $1$, the actual speed of sound $c^2_i=\partial p/\partial \varepsilon$ may exceed the speed of light. To maintain causality, EOS instances with $c^2_s>1$ are excluded from the training database. This differs from the original approach proposed by Fujimoto et al \cite{fujimoto20}.

\begin{table*}[htbp]
\centering
\begin{tabular}{c|c|c} 
\hline \hline
Layer & Number of neurons & Activation function \\\hline
1(Input) & 68 & N/A \\\hline 
2 & 80 & ReLU \\\hline
3 & 60 & ReLU \\\hline
4 & 40 & ReLU \\\hline
5(Output) & 5/6 & tanh \\\hline \hline
\end{tabular}
\caption{
The design structure of DNN in present framework. The measured masses, radii and corresponding variances of $17$ neutron star are set up as the $68$ neurons in the input layer. The output part corresponds to the amount of sound velocity in the remaining segments except the phase transition segment.}\label{tab1} 
\end{table*}

The data generation method closely resembles that employed by Fujimoto et al \cite{fujimoto20}, with the only alteration being the inclusion of $17$ neutron stars instead of the preceding $14$. The three additional neutron stars are PSR J0030+0451 \cite{riley2019,miller2019}, PSR J0740+6620 \cite{riley2021,miller2021}, and 3XMM J185246.6+003317 \cite{3XMM2022}.

For a given EOS, the TOV equation, describing the structure of a spherically symmetric star in static gravitational equilibrium, is solved to obtain the corresponding mass-radius ($M$-$R$) relation. Subsequently, $17$ data points are randomly {within the uniform distribution} selected within the [$M_{\odot}$, $M_{\text{max}}$] interval to generate a set of "original" data points. For each combined set of original data points, $100$ sets representing uncertainties in mass and radius, denoted as $\sigma_{M_i}$ and $\sigma_{R_i}$, respectively, are selected from uniform distributions within the intervals [$0$, $M_{\odot}$] and [$0$, $5$ km]. Within each group of $\sigma_{M_i}$ and $\sigma_{R_i}$, $100$ sets of deviations in mass and radius, denoted as $\Delta M_{ij}$ and $\Delta R_{ij}$ respectively, are chosen. Therefore, each set of "original data points" $M_i, R_i$ corresponds to $100\times100$ sets of "true" data points ($M+\Delta M_{ij}$, $\sigma_{M_i}$ ,$R+\Delta R_{ij}$, $\sigma_{R_i}$).

\begin{figure}[htbp]
    \centering
    \includegraphics[width=0.6\linewidth]{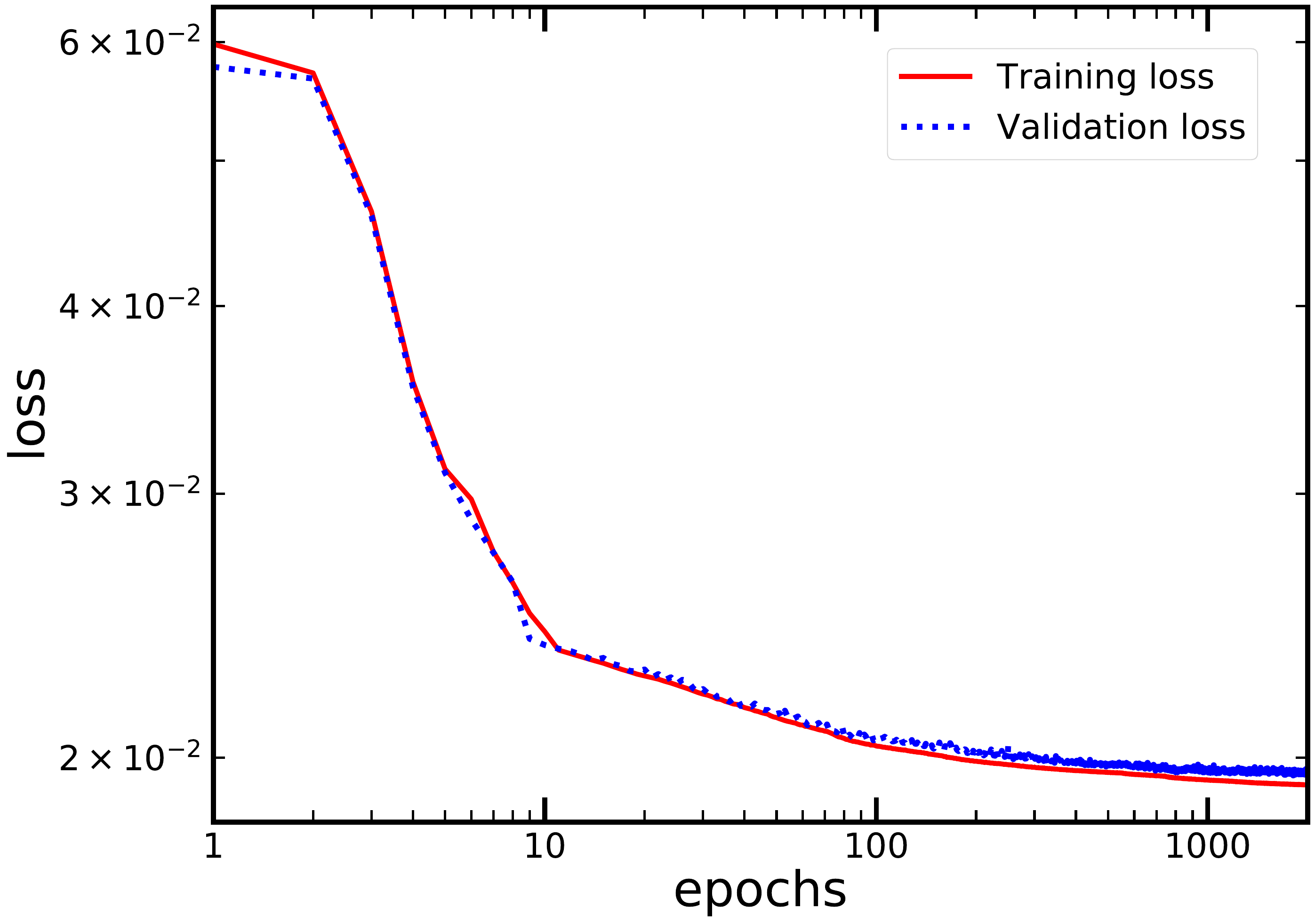}
    \caption{The loss of epochs with the training data and validation data when $\varepsilon_{pt}=2.5\varepsilon_0$ and $\Delta\varepsilon=0.5\varepsilon_0$.}
    \label{fig1}
\end{figure}

The aforementioned process is repeated 500 times to generate a total dataset of $5,000,000$ samples. Among these, $1,000,000$ samples are randomly {within the uniform distribution} selected as training data for the DNN, while $10,000$ samples are chosen as validation data from the remaining $4,000,000$. The DNN architecture is constructed using the Python library Keras \cite{chollet2015} with TensorFlow \cite{abadi2016} as the backend. Specific details of the neural network architecture are provided in Table \ref{tab1}. The loss function 'MSLE', optimization method 'Adam' \cite{kingma2014}, and Glorot Uniform distribution \cite{glorot2010} for parameter initialization are employed, with a batch size of $1000$. To ensure the generation of a robust model, the training epoch of the DNN is tested. As depicted in Fig. \ref{fig1}, the loss of the DNN stabilizes around epoch $100$, indicating adequate training of the model. Therefore, an epoch value of $100$ is selected for all subsequent model training.

\section{Results and discussions}\label{sec3}

In order to enhance the predictive capability of the present DNN framework, an additional constraint is introduced during the dataset generation process. Specifically, it is required that the maximum mass of neutron stars in the training dataset exceeds $2M_{\odot}$. This guarantees the dataset covers enough large-mass neutron stars, hence boosting the DNN's competency in forecasting their attributes. Moreover, the uncertainties associated with the neutron star observables included in the training dataset, denoted as $\Delta M_i$ and $\Delta R_i$, are incorporated into the DNN parameters, introducing fluctuations in the prediction results. To quantify this uncertainty, 100 independent sets of DNNs are trained for each phase transition scenario, characterized by a specific combination of $(\varepsilon_{pt},~\Delta \varepsilon)$. It is assumed that the predicted EOSs generated by the 100 sets of DNNs follow a Gaussian distribution. Subsequently, Gaussian fitting is employed to infer the possible average speeds of sound, $\langle c^2_{s,i}\rangle$, at the energy density segments without phase transitions. The uncertainties associated with these inferred average speeds of sound are calculated at $1\sigma~(68.3\%)$ or $2\sigma ~(95.4\%)$ confidence levels.

\begin{figure*}[htb]
	\centering
	\includegraphics[width=0.8\textwidth]{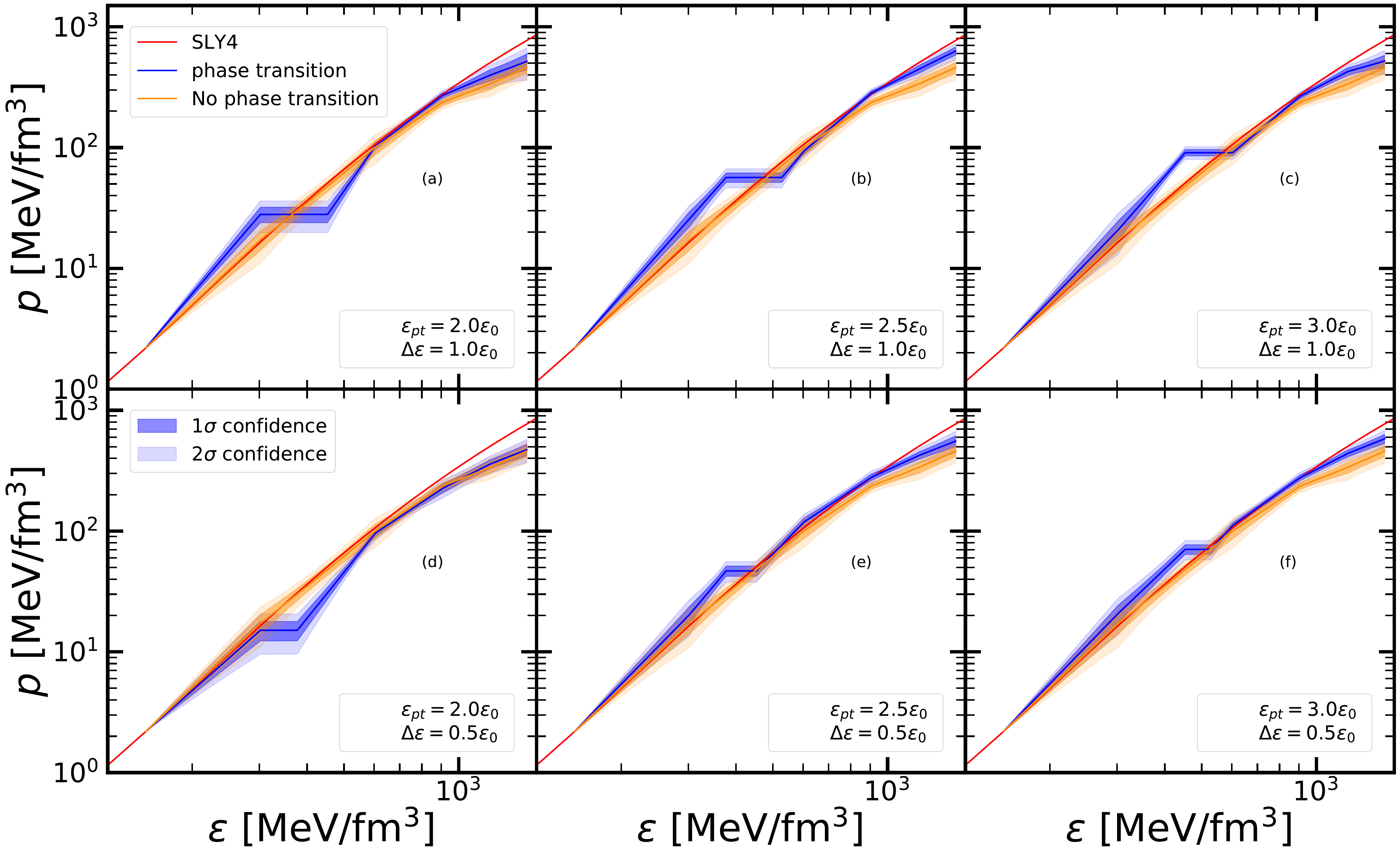}
	\caption{The EOSs deduced from the DNN framework with different starting points, $\varepsilon_{pt}$ and lengths $\Delta\varepsilon$ of phase transition and compared with the EOSs without phase transition from DNN and SLy4 set. }
	\label{fig2}
\end{figure*}

In Fig. \ref{fig2}, the EOSs obtained from the DNNs with $1\sigma$ and $2\sigma$ confidence levels, considering phase transitions using the Maxwell construction, are displayed and compared with those not incorporating phase transitions ({It means where the speed of sound of neutron matter should be larger than zero in these EOS and the first-order phase transition with Maxwell construction does not happen.}), as well as the EOS deduced from the Skyrme-Hartree-Fock model (SLy4 set) for hadronic matter \cite{sly4}. It is evident that the EOSs generated without phase transitions closely match those derived from SLy4 below $\varepsilon=1000$ MeV/fm$^3$. In our current computations, we explore two different lengths of phase transition segments, $\Delta\varepsilon = 0.5\varepsilon_0$ and $\Delta\varepsilon = 1.0\varepsilon_0$, and contemplate three onset energy densities for the phase transition, $\varepsilon_{pt}=2.0,~2.5,~3.0\varepsilon_0$, predicated on available  researches on phase transitions in neutron stars.

As exhibited in panel (a), when the length of the phase transition region is larger, $\Delta\varepsilon=1.0\varepsilon_0$, and its onset density is smaller, $\varepsilon_{pt}=2.0\varepsilon_0$, the EOS before the occurrence of the phase transition becomes stiffer in comparison to the one devoid of the phase transition. As it is a first-order phase transition, the pressure remains constant at the energy density of the phase transition segment. Once nucleons deconfine into quarks, the pressure increases rapidly and becomes larger than the pressure without the phase transition, leading to the formation of massive neutron stars. Conversely, if the phase transition occurs at a higher density, such as $\varepsilon_{pt}=3.0\varepsilon_0$ as shown in panel (c), the EOSs before the phase transition are still stiffer compared to those of pure hadronic matter, but they closely resemble the EOSs without phase transition at the end of the phase transition segment. The increase in pressure is gradual compared to the scenario in panel (a). Pure quark matter emerges at $\varepsilon=4.0\varepsilon_0$, closer to the core of the neutron star, and exhibits a higher stiffness, resulting in the generation of heavier stars. Similar situations arise in cases with higher onset energy densities, as depicted in panels (e) and (f).

The behaviors of EOS before the phase transition, denoted by smaller values of $\Delta\varepsilon$ and $\varepsilon_{pt}$ as depicted in panel (d), exhibits marked distinctions. It resembles typical behavior of normal hadronic matter underway to the phase transition. The EOS rapidly increases towards the end of the phase transition but does not transition to a stiffer EOS. Ultimately, it represents the softest EOS among the six cases considered.

{Generally speaking, a softer EOS is preferred in the low mass region due to the $M$-$R$ observations. To support heavy neutron stars, however, the EOS should become stiffer after the phase transition. When the pressure increases too rapidly in high-density regions, the real speed of sound may exceed that of light, especially in the case of a first-order phase transition compared to one without a phase transition. Furthermore, the mass input of 100 deep neural networks is uniformly chosen from the observables. Consequently, the present EOSs are biased towards intermediate mass neutron stars, resulting in a less stiff EOS in high-density regions compared to the SLy4 case for pure hadronic matter. In the future, if more data on heavy neutron stars becomes available, the EOS is expected to become generally stiffer. In our previous work discussing phase transitions in neutron stars \cite{ju2021b}, we also found that the EOS for quark matter becomes softer in high-density regions compared to the EOS for pure hadronic matter, after considering the Maxwell construction at a first-order phase transition. This is consistent with the present results from machine learning.
}

\begin{figure*}[htbp]
	\centering
	\includegraphics[width=0.8\textwidth]{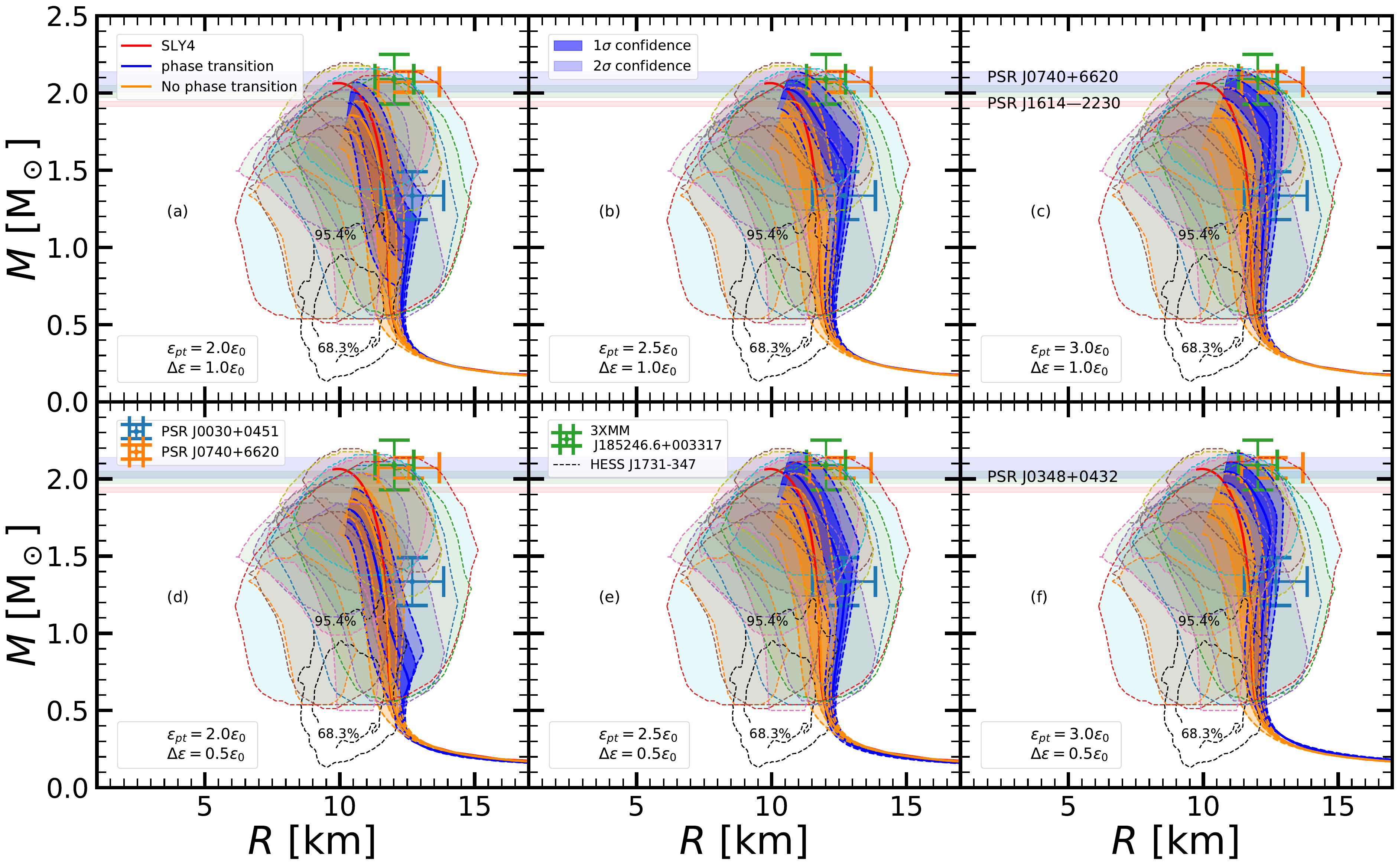}
	\caption{$M$-$R$ relations corresponding to the deduced EOSs from DNNs with and without phase transitions in the $1\sigma$ and $2\sigma$ confidences. The mass-radius observation distributions from $14$ neutron stars, and the mass-radius error bars from new added 3 neutron stars are compared to. The black dashed line is the observables of the central compact object at HESS J1731-347.}
	\label{fig3}
\end{figure*}

The mass-radius ($M$-$R$) relations of neutron stars are derived by solving the TOV equation once the EOSs are inferred from the DNNs. These relations are depicted in Fig. \ref{fig3} with shaded regions representing $1\sigma$ confidence (darker shade) and $2\sigma$ confidence (lighter shade). Concurrently, the observational data on the mass-radius distribution for $14$ neutron stars are presented with contour lines, while data from an additional $3$ neutron stars are displayed with error bars. The results obtained from the EOSs without phase transition, generated by the DNNs, as well as the SLy4 set, are also included for comparison. In general, the radii of neutron stars encompassing phase transitions tend to be larger in certain mass regions compared to those derived from pure hadronic matter. As the onset energy density of the quark phase increases, the leap in radius occurs at larger masses. Furthermore, the maximum masses of neutron stars involving phase transitions are greater than those without phase transitions and align more closely with the constraints from NICER measurements for PSR J0030+0451 and PSR J0740+6620 when $\varepsilon\geq 2.5 \varepsilon_0$.

Additionally, we investigate whether our EOSs inferred from DNNs, that exclude low-mass neutron stars in the training set, can satisfy recent constraints on such objects. The black dashed and solid lines in Fig. \ref{fig3} represent the $68\%$ and $95\%$ confidence regions, respectively, of the central compact object of HESS J1731-347. Its mass and radius are reported as $M=0.77^{+0.20}_{-0.17} M_\odot$ and $R=10.4^{+0.86}_{-0.78}$ km \cite{doroshenko2022}. It is evident that the $M$-$R$ relations, including the phase transition, can only describe the $95.4\%$ confidence interval of HESS J1731-347. Should its radius be smaller, it suggests the absence of a phase transition in this compact object. According to available investigations, the phase transition is expected to occur in neutron stars with masses above $1.5 M_\odot$.

\begin{figure*}[htbp]
    \centering
    \includegraphics[width=0.8\textwidth]{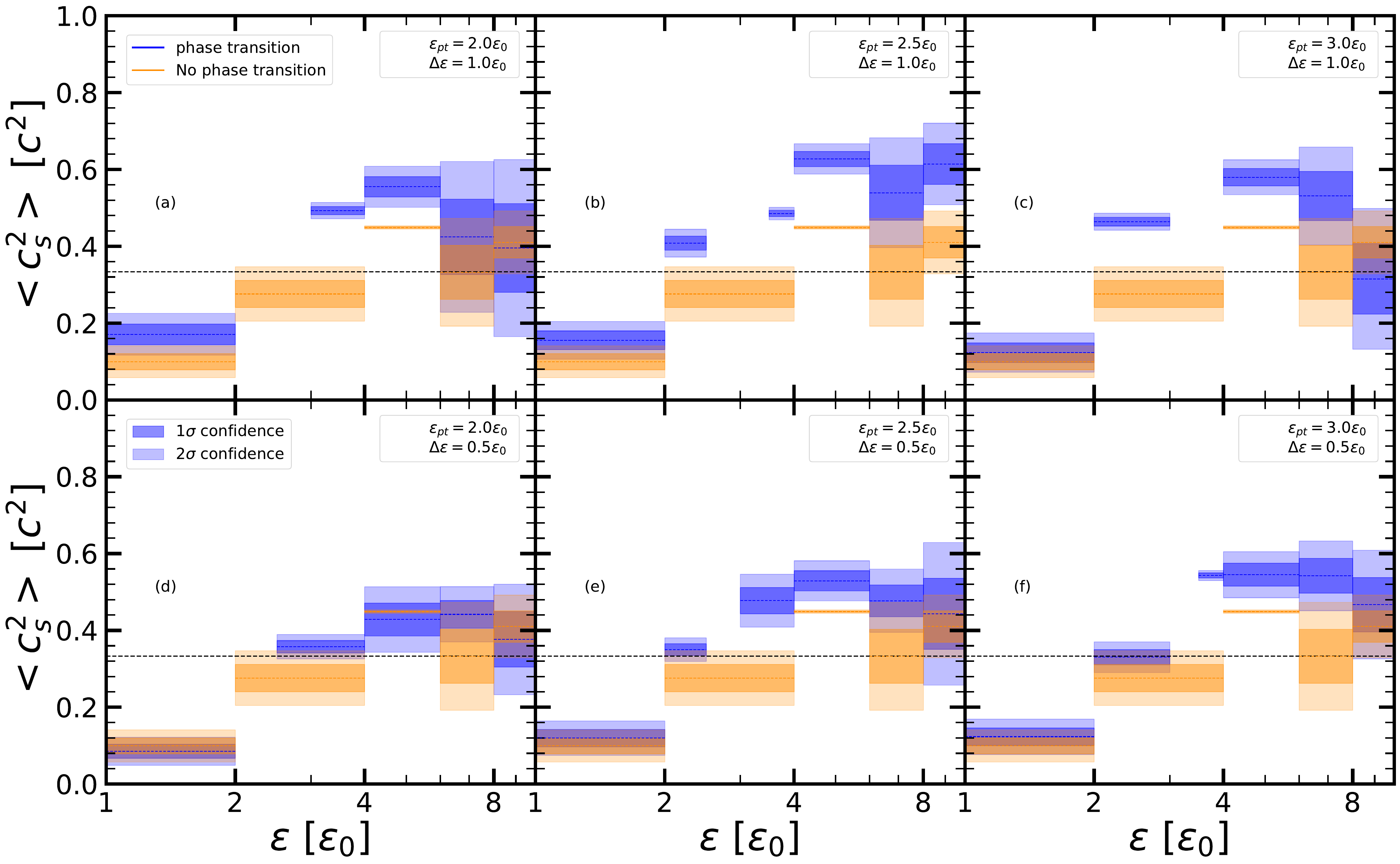}
    \caption{
    Th average speeds of sound, $\langle c^2_{s,i}\rangle$, generated by the DNNs with and without phase transition for different $\varepsilon_{pt}$ and $\Delta\varepsilon$. The horizontal dotted line represents the conformal limit of $\langle c_s^2\rangle=1/3$.
    }
    \label{fig4}
\end{figure*}

In Fig. \ref{fig4}, the average speed of sound of neutron star matter, $\langle c^2_s\rangle$, at different energy density segments trained by the DNNs is plotted for cases both with and without phase transition. Without phase transition, the average speed of sound increases up to around $6\varepsilon_0$ and then decreases between $6\varepsilon_0$ and $8\varepsilon_0$, approaching the conformal limit $c^2_s=1/3$ at high density regions. Throughout the entire density range, its magnitude remains below $0.5$. Conversely, with phase transition, the average speed of sound tends to be generally larger than that without phase transition, resulting in a stiffer EOS. The maximum value of $\langle c^2_s\rangle$, including quark matter, reaches around $0.7$.

\begin{figure*}[htbp]
    \centering
    \includegraphics[width=0.8\textwidth]{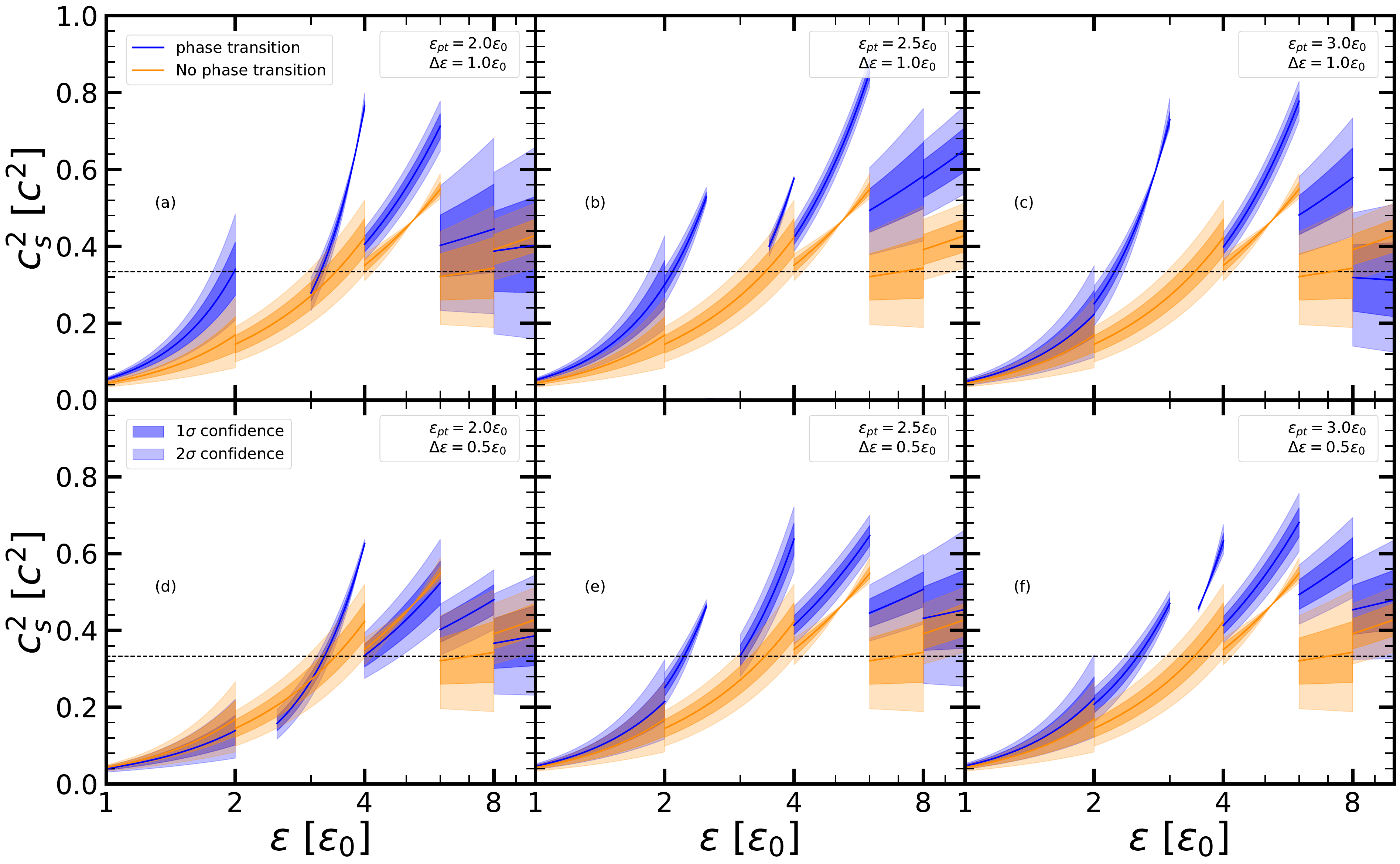}
    \caption{
    The actual speeds of sound obtained from the DNNs with and without phase transition for different $\varepsilon_{pt}$ and $\Delta \varepsilon$.
    }
    \label{fig5}
\end{figure*}

To ensure causality in neutron stars, it's crucial to verify that the speed of sound throughout the EOS remains below the speed of light. While the average speed of sound, $\langle c^2_s\rangle$, may be less than $1$, it's important to examine the actual speed of sound, $c^2_s = \partial p / \partial \varepsilon$, within each energy density segment for the piecewise-polytropic representation. In Fig. \ref{fig5}, we observe that the speed of sound at the end of each segment is notably higher than the average sound speed. Some endpoint speeds even approach $0.9c^2$, but all remain below the speed of light. This can be attributed to an extra constraint incorporated during the training process in our framework, which ascertains $\partial p/\partial \varepsilon \leq 1$. This differs from the original scheme proposed by Fujimoto et al \cite{fujimoto20,fujimoto21}.

\begin{figure*}[htbp]
    \centering
    \includegraphics[width=0.8\textwidth]{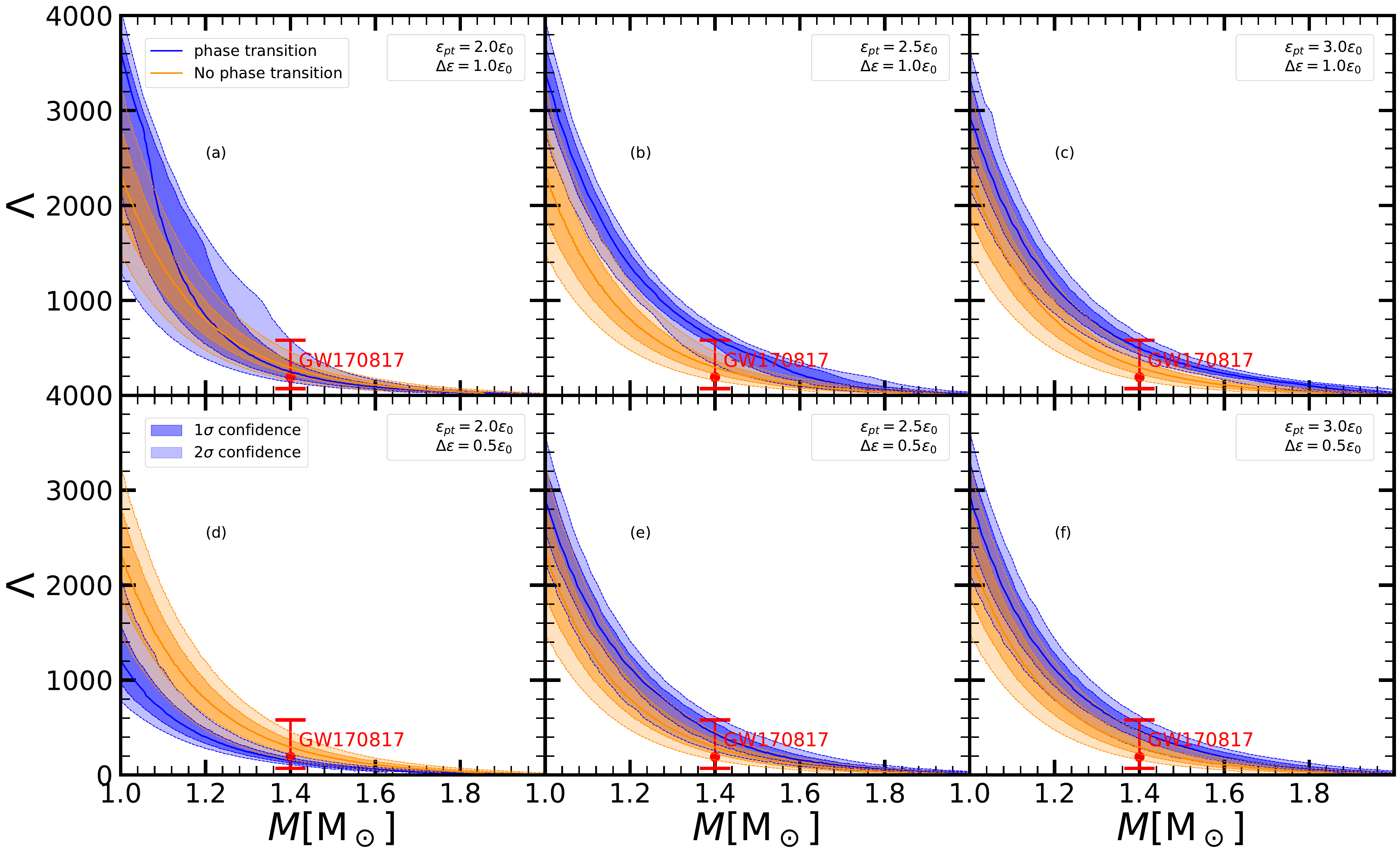}
    \caption{Tidal deformabilities $\Lambda$ as functions of neutrons star masses generated by the DNNs with and without phase transition and comparing to the values extracted from GW170817 event (red bar).}
    \label{fig6}
\end{figure*}

The tidal deformability has emerged as another crucial global property of neutron stars, particularly since the detection of the gravitational wave event GW170817 by the LIGO and Virgo collaborations in 2017. This event, resulting from the merger of the first binary neutron star system, provided valuable insights into the quadrupole deformation of compact stars within the gravitational field of their companions, with the tidal deformability strongly linked to the EOS of neutron star matter. The dimensionless tidal deformability, $\Lambda_{1.4}$, was inferred as $190^{+390}_{-120}$ at $1.4M_\odot$ from the GW170817 event. Our calculations of tidal deformabilities from the DNNs, encompassing both with and without phase transition, are depicted in Fig. \ref{fig6}, and are compared to the constraints derived from gravitational wave measurements.

Concerning pure hadronic matter, the dimensionless tidal deformabilities obtained from our DNNs are $\Lambda_{1.4}=294.37^{+77.92}_{-69.75}$ and $\Lambda_{1.4}=294.37^{+164.50}_{-134.20}$ at $68\%$ and $95\%$ confidence levels, respectively. These results are in complete agreement with the constraints derived from GW170817. Fujimoto et al. made a similar prediction of $\Lambda_{1.4}=320^{+120}_{-120}$ using a comparable framework \cite{fujimoto20}. If the phase transition appears earlier, such as $\varepsilon_{pt}=2.0\varepsilon$,  $\Lambda_{1.4}$ will become smaller comparing to the case without the phase transition. In particular, it is just  $151.52^{+30.30}_{-23.57}$ for the short phase transition interval $\Delta\varepsilon=0.5\varepsilon_0$ at $68\%$ confidence level, since the radius of the neutron star is obviously reduced as $10.70$ km. The dimensionless tidal deformabilities increases and are about $430$ to $600$ at $1.4M_\odot$ when the onset density of quark matter is larger than $2.5\varepsilon_0$. The bigger phase transition interval $\Delta\varepsilon$ also can generate the larger value of $\Lambda_{1.4}$ due to growth of radius.

In Tables \ref{tab2} and \ref{tab3}, we provide detailed properties of neutron stars, including the density jump, $\eta$, the maximum mass, $M_{\max}$, the radius of the maximum mass, $R_{\max}$, the radius at $1.4 M_\odot$, $R_{1.4}$, the dimensionless tidal deformability at $1.4 M_\odot$, $\Lambda_{1.4}$, and the mass, radius, and density at the appearance of the phase transition, $M_{\text{trans}}$, $R_{\text{trans}}$, and $\rho_{\text{trans}}$, for $\Delta\varepsilon=1.0\varepsilon_0$ and $\Delta\varepsilon=0.5\varepsilon_0$.

\begin{table}[htbp]
	\begin{tabular}{c|ccccccccc}
		\hline
		$\varepsilon_{pt}$      & C.L.                    & $\eta$                                     & $M_{\text{max}}${[}M$_\odot${]}                   & $R_{\text{max}}${[}km{]}                        & $R_{1.4}${[}km{]}                        & $\Lambda_{1.4}$                            & $M_{\text{trans}}${[}M$_\odot${]}                       & $R_{\text{trans}}${[}km{]}                            & $\rho_{\text{trans}}${[}fm$^{-3}${]}                       \\ \hline
		\multirow{4}{*}{2.0} & \multirow{2}{*}{$68\%$} & \multirow{2}{*}{$0.458^{+0.006}_{-0.006}$} & \multirow{2}{*}{$1.936^{+0.071}_{-0.078}$} & \multirow{2}{*}{$10.34^{+0.11}_{-0.05}$} & \multirow{2}{*}{$11.59^{+0.52}_{-0.45}$} & \multirow{2}{*}{$251.08^{+103.90}_{-69.26}$}  & \multirow{2}{*}{$1.170^{+0.150}_{-0.158}$} & \multirow{2}{*}{$12.05^{+0.27}_{-0.27}$} & \multirow{2}{*}{$0.314^{+0.001}_{-0.001}$} \\
		&   &   &   &   &   &   &   &  \\\cline{2-10} 
		& \multirow{2}{*}{$95\%$} & \multirow{2}{*}{$0.458^{+0.012}_{-0.011}$} & \multirow{2}{*}{$1.936^{+0.136}_{-0.164}$} & \multirow{2}{*}{$10.34^{+0.19}_{-0.14}$} & \multirow{2}{*}{$11.59^{+1.19}_{-0.84}$} & \multirow{2}{*}{$251.08^{+333.33}_{-112.55}$} & \multirow{2}{*}{$1.170^{+0.291}_{-0.325}$} & \multirow{2}{*}{$12.05^{+0.52}_{-0.56}$} & \multirow{2}{*}{$0.314^{+0.002}_{-0.001}$} \\
		&   &   &   &   &   &   &   &  \\\hline 
		\multirow{4}{*}{2.5} & \multirow{2}{*}{$68\%$} & \multirow{2}{*}{$0.347^{+0.004}_{-0.004}$} & \multirow{2}{*}{$2.027^{+0.058}_{-0.061}$} & \multirow{2}{*}{$10.52^{+0.18}_{-0.14}$} & \multirow{2}{*}{$12.69^{+0.18}_{-0.25}$} & \multirow{2}{*}{$597.40^{+64.94}_{-73.59}$}   & \multirow{2}{*}{$1.642^{+0.112}_{-0.119}$} & \multirow{2}{*}{$12.33^{+0.24}_{-0.26}$} & \multirow{2}{*}{$0.385^{+0.002}_{-0.002}$} \\
		&   &   &   &   &   &   &   &  \\\cline{2-10} 
		& \multirow{2}{*}{$95\%$} & \multirow{2}{*}{$0.347^{+0.008}_{-0.008}$} & \multirow{2}{*}{$2.027^{+0.112}_{-0.126}$} & \multirow{2}{*}{$10.52^{+0.33}_{-0.29}$} & \multirow{2}{*}{$12.69^{+0.36}_{-0.90}$} & \multirow{2}{*}{$597.40^{+125.54}_{-274.65}$} & \multirow{2}{*}{$1.642^{+0.220}_{-0.243}$} & \multirow{2}{*}{$12.33^{+0.46}_{-0.54}$} & \multirow{2}{*}{$0.385^{+0.003}_{-0.003}$} \\
		&   &   &   &   &   &   &   &  \\\hline 
		\multirow{4}{*}{3.0} & \multirow{2}{*}{$68\%$} & \multirow{2}{*}{$0.278^{+0.003}_{-0.003}$} & \multirow{2}{*}{$2.032^{+0.060}_{-0.064}$} & \multirow{2}{*}{$10.88^{+0.16}_{-0.12}$} & \multirow{2}{*}{$12.37^{+0.24}_{-0.26}$} & \multirow{2}{*}{$489.18^{+73.59}_{-64.94}$}   & \multirow{2}{*}{$1.804^{+0.106}_{-0.112}$} & \multirow{2}{*}{$12.37^{+0.21}_{-0.22}$} & \multirow{2}{*}{$0.453^{+0.003}_{-0.003}$} \\
		&   &   &   &   &   &   &   &  \\\cline{2-10} 
		& \multirow{2}{*}{$95\%$} & \multirow{2}{*}{$0.278^{+0.006}_{-0.006}$} & \multirow{2}{*}{$2.032^{+0.117}_{-0.132}$} & \multirow{2}{*}{$10.88^{+0.30}_{-0.24}$} & \multirow{2}{*}{$12.37^{+0.47}_{-0.53}$} & \multirow{2}{*}{$489.18^{+160.17}_{-108.23}$} & \multirow{2}{*}{$1.804^{+0.204}_{-0.233}$} & \multirow{2}{*}{$12.37^{+0.38}_{-0.47}$} & \multirow{2}{*}{$0.453^{+0.006}_{-0.005}$} \\
		&   &   &   &   &   &   &   &  \\\hline                             
	\end{tabular}
	\caption{The properties of neutrons star from the DNNs with different onset energy densities of phase transition for $\Delta\varepsilon = 1.0\varepsilon_0$ at $68\%$ and $95\%$ confidence levels, respectively.}	\label{tab2}
\end{table}

The density jump, $\eta$, is defined as $\eta \equiv \frac{\rho_{-}}{\rho_{+}} - 1$, where $\rho_{-}$ ($\rho_{+}$) is the density at the top of the quark phase (bottom of the hadronic phase) in the Maxwell construction during a first-order phase transition. A larger $\eta$ indicates a stronger phase transition, which corresponds to smaller $\varepsilon_{\text{pt}}$ and larger $\Delta\varepsilon$. Lima et al. estimated $\eta=0.129^{+0.211}_{-0.099}$ and $M_{\text{trans}}=1.39^{+0.36}_{-0.31}M_{\odot}$ with a $1\sigma$ confidence level from 3XMM J1852+0033, and $\eta=0.243^{+0.377}_{-0.183}$ and $M_{\text{trans}}=1.50^{+0.29}_{-0.26}M_{\odot}$ from PSR J0030+0451 \cite{3XMM2022}.

The density jumps in this study, with $\varepsilon_{pt}$ ranging from $2.0\varepsilon_0$ to $3.0\varepsilon_0$, and $\Delta\varepsilon$ values of $0.5\varepsilon_0$ and $1.0\varepsilon_0$, range from approximately $0.145$ to $0.458$, consistent with estimates from observations of 3XMM J1852+0033 and PSR J0030+0451. The uncertainties in density jump and $\rho_\text{trans}$ are introduced by the following expression: $d\varepsilon/(\varepsilon+p)=d\rho/\rho$. The pressures have different values at the onset energy density $\varepsilon_{pt}$ from the $100$ DNNs. The maximum masses of neutron stars with a phase transition density above $2.5\varepsilon_0$ are around $2.0 M_\odot$, whereas they are only $1.8$ to $1.9M_\odot$ when quarks emerge at $2.0\varepsilon_0$. In the former case, the radius and deformability at $1.4M_\odot$ are relatively larger compared to the latter case. Furthermore, the neutron star mass, radius, and density at the beginning of the phase transition also noticeably increase. The phase transition can occur around $1.0 M_\odot$ at $\varepsilon_{pt}=2.0\varepsilon_0$.

\begin{table}[htbp]
\begin{tabular}{c|ccccccccc}
\hline
$\varepsilon_{pt}$      & C.L.                    & $\eta$                                     & $M_{\text{max}}${[}M$_\odot${]}                   & $R_{\text{max}}${[}km{]}                        & $R_{1.4}${[}km{]}                        & $\Lambda_{1.4}$                             & $M_{\text{trans}}${[}M$_\odot${]}                       & $R_{\text{trans}}${[}km{]}                            & $\rho_{\text{trans}}${[}fm$^{-3}${]}                       \\ \hline
\multirow{4}{*}{2.0} & \multirow{2}{*}{$68\%$} & \multirow{2}{*}{$0.240^{+0.002}_{-0.002}$} & \multirow{2}{*}{$1.802^{+0.073}_{-0.080}$} & \multirow{2}{*}{$9.90^{+0.11}_{-0.12}$}  & \multirow{2}{*}{$10.70^{+0.30}_{-0.28}$} & \multirow{2}{*}{$151.52^{+30.30}_{-23.57}$}   & \multirow{2}{*}{$0.933^{+0.113}_{-0.115}$} & \multirow{2}{*}{$10.92^{+0.31}_{-0.32}$} & \multirow{2}{*}{$0.317^{+0.001}_{-0.001}$} \\
                     &   &   &   &   &   &   &   &  \\\cline{2-10} 
                     & \multirow{2}{*}{$95\%$} & \multirow{2}{*}{$0.240^{+0.004}_{-0.004}$} & \multirow{2}{*}{$1.802^{+0.140}_{-0.168}$} & \multirow{2}{*}{$9.90^{+0.21}_{-0.25}$}  & \multirow{2}{*}{$10.70^{+0.60}_{-0.57}$} & \multirow{2}{*}{$151.52^{+64.94}_{-43.29}$}   & \multirow{2}{*}{$0.933^{+0.222}_{-0.233}$} & \multirow{2}{*}{$10.92^{+0.59}_{-0.68}$} & \multirow{2}{*}{$0.317^{+0.001}_{-0.001}$} \\
                     &   &   &   &   &   &   &   &  \\\hline 
\multirow{4}{*}{2.5} & \multirow{2}{*}{$68\%$} & \multirow{2}{*}{$0.176^{+0.002}_{-0.002}$} & \multirow{2}{*}{$2.040^{+0.069}_{-0.074}$} & \multirow{2}{*}{$10.73^{+0.20}_{-0.17}$} & \multirow{2}{*}{$12.19^{+0.36}_{-0.41}$} & \multirow{2}{*}{$437.23^{+108.23}_{-103.90}$} & \multirow{2}{*}{$1.543^{+0.125}_{-0.133}$} & \multirow{2}{*}{$12.03^{+0.25}_{-0.26}$} & \multirow{2}{*}{$0.388^{+0.002}_{-0.002}$} \\
                     &   &   &   &   &   &   &   &  \\\cline{2-10} 
                     & \multirow{2}{*}{$95\%$} & \multirow{2}{*}{$0.176^{+0.004}_{-0.004}$} & \multirow{2}{*}{$2.040^{+0.135}_{-0.153}$} & \multirow{2}{*}{$10.73^{+0.38}_{-0.33}$} & \multirow{2}{*}{$12.19^{+0.56}_{-0.83}$} & \multirow{2}{*}{$437.23^{+160.17}_{-190.48}$} & \multirow{2}{*}{$1.543^{+0.242}_{-0.274}$} & \multirow{2}{*}{$12.03^{+0.47}_{-0.57}$} & \multirow{2}{*}{$0.388^{+0.003}_{-0.003}$} \\
                     &   &   &   &   &   &   &   &  \\\hline 
\multirow{4}{*}{3.0} & \multirow{2}{*}{$68\%$} & \multirow{2}{*}{$0.145^{+0.002}_{-0.002}$} & \multirow{2}{*}{$2.046^{+0.063}_{-0.066}$} & \multirow{2}{*}{$10.68^{+0.21}_{-0.17}$} & \multirow{2}{*}{$12.26^{+0.28}_{-0.31}$} & \multirow{2}{*}{$463.20^{+77.92}_{-82.25}$}   & \multirow{2}{*}{$1.726^{+0.113}_{-0.120}$} & \multirow{2}{*}{$11.98^{+0.24}_{-0.27}$} & \multirow{2}{*}{$0.456^{+0.003}_{-0.003}$} \\
                     &   &   &   &   &   &   &   &  \\\cline{2-10} 
                     & \multirow{2}{*}{$95\%$} & \multirow{2}{*}{$0.145^{+0.004}_{-0.004}$} & \multirow{2}{*}{$2.046^{+0.123}_{-0.136}$} & \multirow{2}{*}{$10.68^{+0.39}_{-0.35}$} & \multirow{2}{*}{$12.26^{+0.52}_{-0.78}$} & \multirow{2}{*}{$463.20^{+147.19}_{-177.49}$} & \multirow{2}{*}{$1.726^{+0.220}_{-0.250}$} & \multirow{2}{*}{$11.98^{+0.46}_{-0.56}$} & \multirow{2}{*}{$0.456^{+0.006}_{-0.005}$} \\
                     &   &   &   &   &   &   &   &  \\\hline                      
\end{tabular}
\caption{The properties of neutrons star from the DNNs with different onset energy densities of phase transition for $\Delta\varepsilon = 0.5\varepsilon_0$ at $68\%$ and $95\%$ confidence levels, respectively.}	\label{tab3}
\end{table}

\begin{figure}[htbp]
	\centering
	\includegraphics[width=0.8\textwidth]{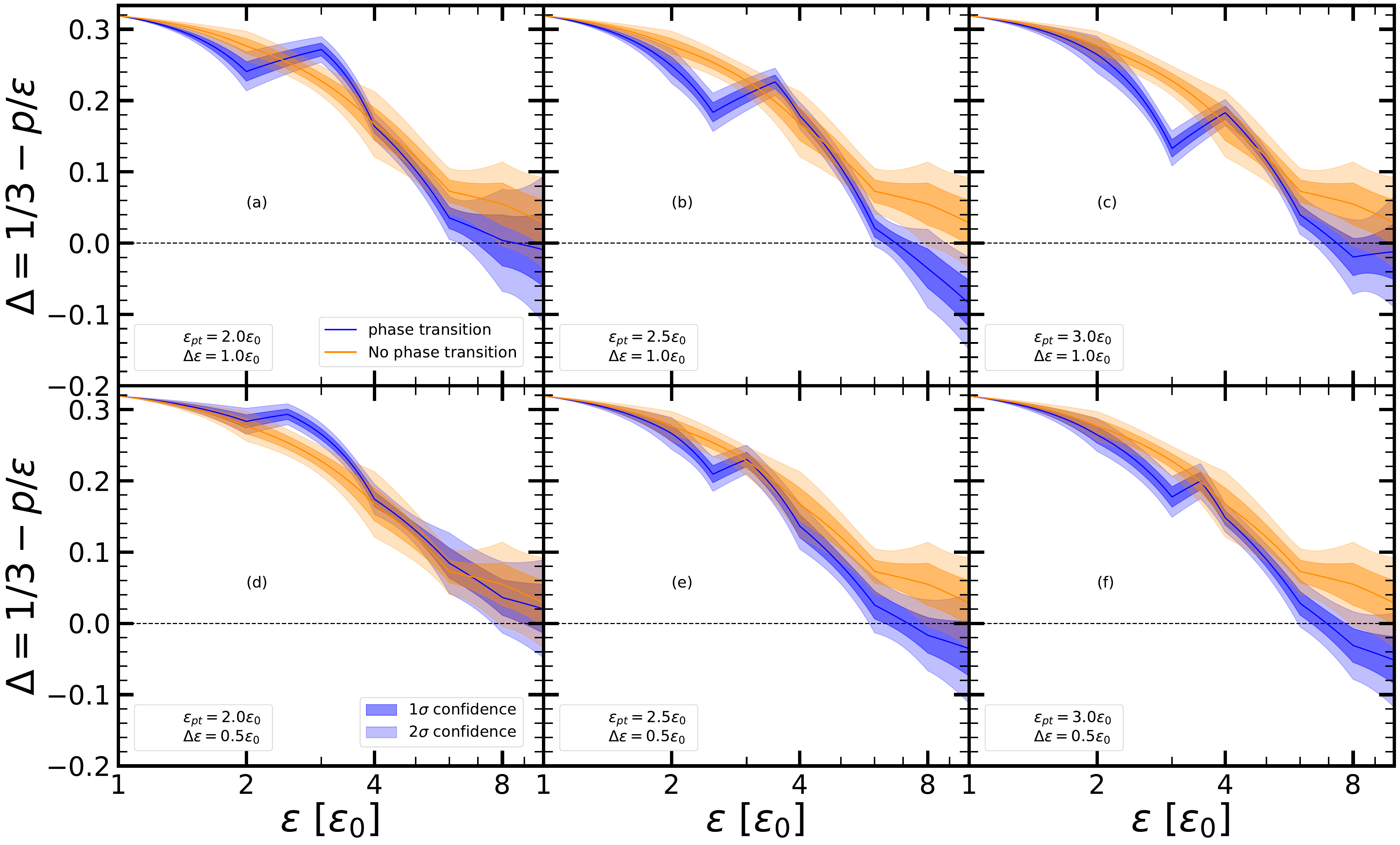}
	\caption{The trace anomaly as a function of energy density from present DNNs with and without phase transition. }
	\label{fig7}
\end{figure}

Recently, Fujimoto et al. defined the limits of the trace anomaly as:
\begin{equation}
	\Delta = \frac{1}{3} - \frac{p}{\varepsilon},
\end{equation}
which is proposed as a scale of conformality \cite{fujimoto22}. With the requirements of thermodynamic stability and causality, $\Delta$ should satisfy $-2/3 < \Delta \leq 1/3$. It will approach zero at the conformal limit. In Fig. \ref{fig7}, the trace anomalies with and without phase transitions for different $\varepsilon_{pt}$ and $\Delta\varepsilon$ from the DNNs are shown, respectively. It smoothly decreases from $1/3$ to $0$ for pure hadronic matter. When the quark matter appears, the trace anomaly will rapidly increase since pressures remain constant with energy density increment at the phase transition segment. It also decreases with density after the end of the phase transition. At $6\varepsilon_0$, it is less than zero, which is consistent with recent results by Tak\'asty et al \cite{takatsy2023}. Above this energy density, the trace anomalies from the phase transition are lower than those without phase transition for $\varepsilon_{pt} \geq 2.5\varepsilon_0$.

\begin{figure}[htbp]
    \centering
    \includegraphics[width=0.4\textwidth]{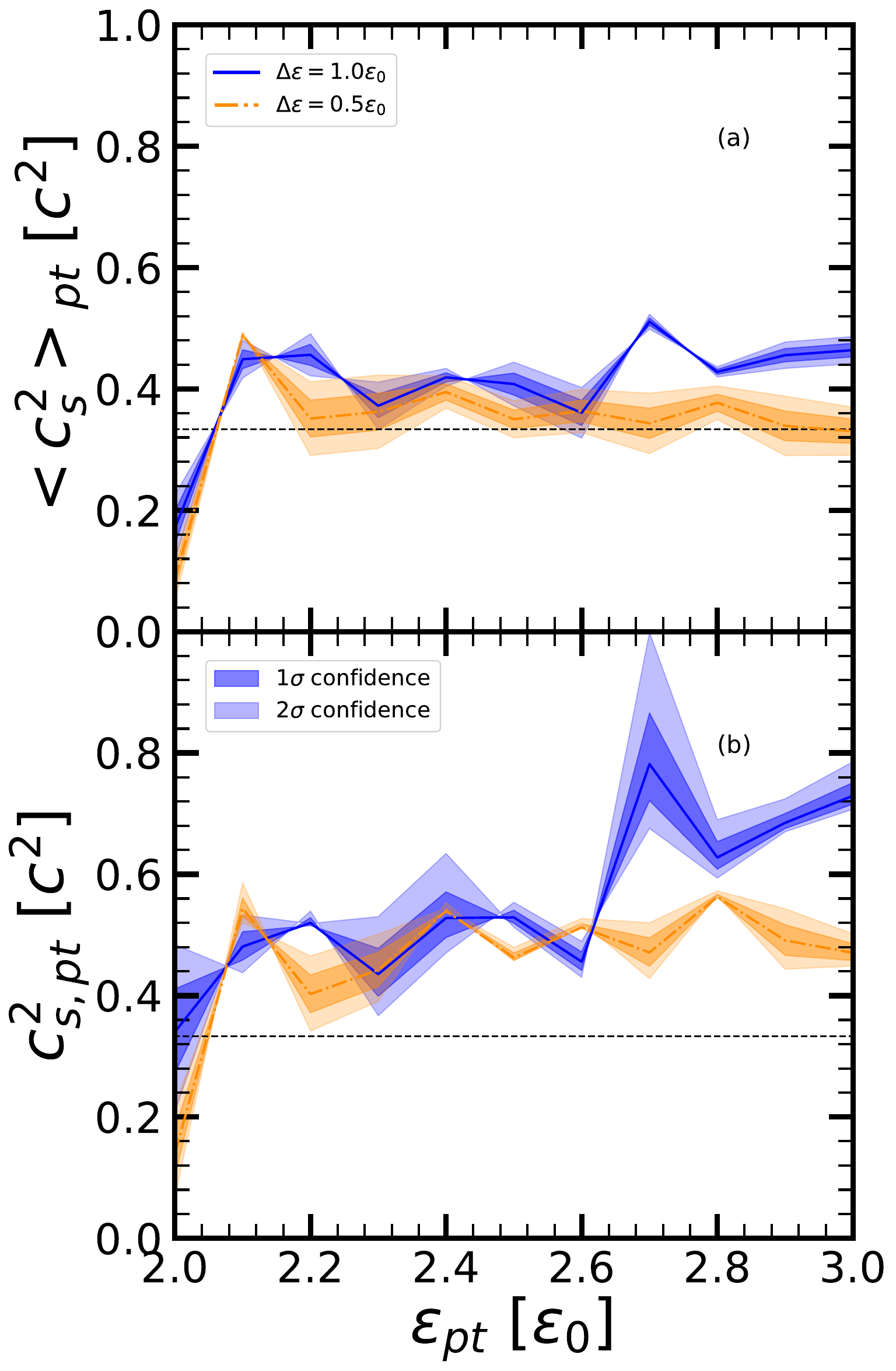}
    \caption{
    The average speed (panel (a)) and real speed (panel (b)) of sound $\langle c_s^2 \rangle$ at the onset energy density of phase transition with respect to $\varepsilon_{pt}$, for $\Delta\varepsilon=1.0\varepsilon_0$ and $\Delta\varepsilon=0.5\varepsilon_0$.}
    \label{fig8}
\end{figure}

To further investigate the properties of phase transitions, Fig. \ref{fig8} depicts the average speeds $\langle c^2_s\rangle_{pt}$ (panel (a)) and the real speeds $c^2_{s,pt}$ (panel (b)) of sound at the onset energy density with different $\varepsilon_{pt}$ ranging from $2\varepsilon_0$ to $3\varepsilon_0$. For average speeds, they range from around $1/3$ to $0.5$ when the phase transition segment is larger with $\Delta\varepsilon=1.0\varepsilon_0$. As this segment decreases to $\Delta\varepsilon=0.5\varepsilon_0$, the average speed $\langle c^2_s\rangle_{pt}$ approaches the conformal limit. Naturally, $\langle c^2_s\rangle_{pt}$ at $2\varepsilon_0$ should be the same as that within the range $[\varepsilon_0, 2\varepsilon_0]$. The real speeds of sound $c^2_{s,pt}$ range from approximately $0.4$ to $0.6$ when the phase transition occurs below $2.6\varepsilon_0$. They exceed $0.6$ when the phase transition occurs later and $\Delta\varepsilon=1.0\varepsilon_0$.

\begin{figure}[htbp]
	\centering
	\includegraphics[width=0.8\textwidth]{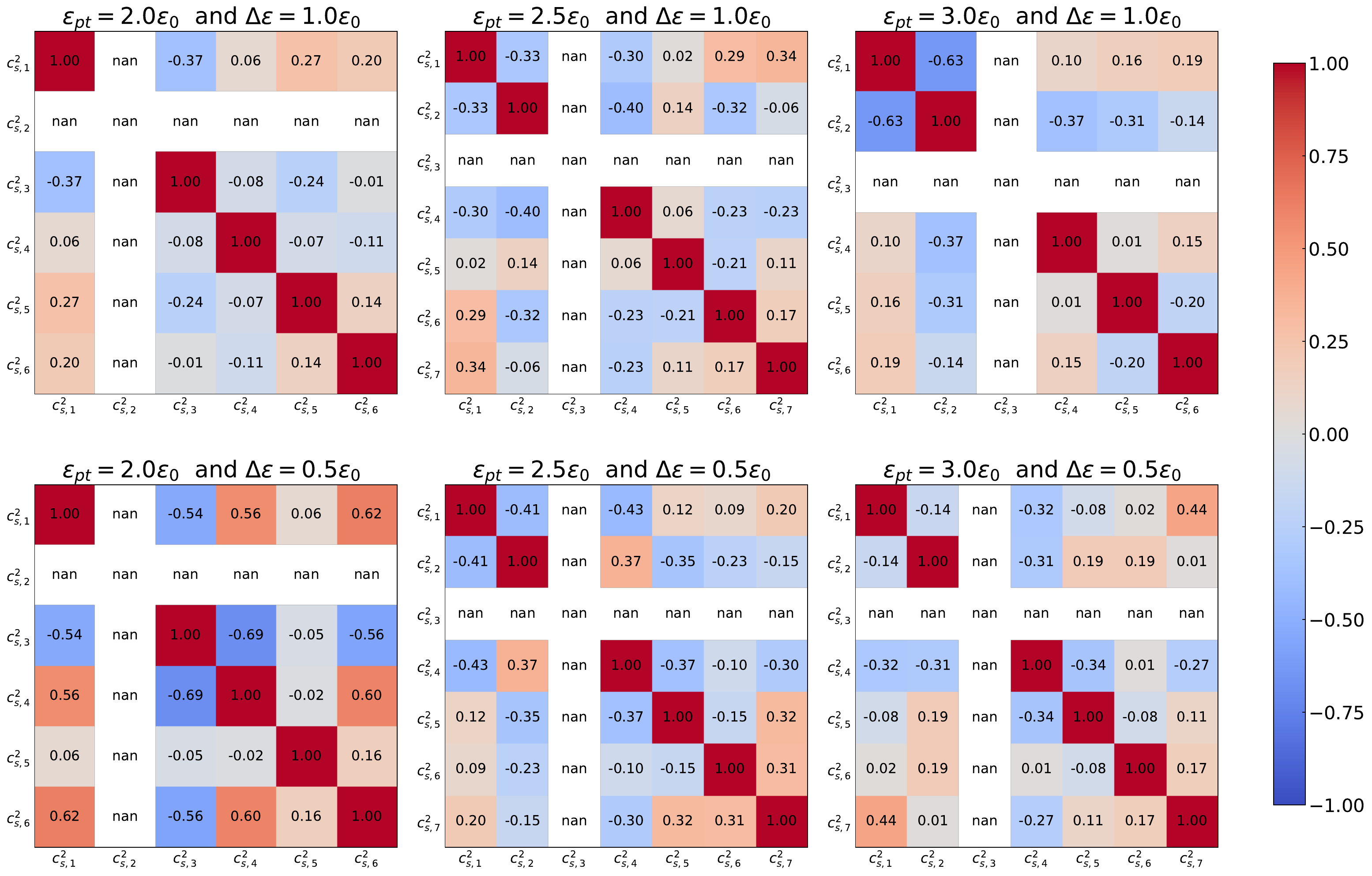}
	\caption{ The correlations of average speed, $\langle c^2_{s,i}\rangle$ at different segments from DNNs at distinguished phase transition conditions.}
	\label{fig9}
\end{figure}

For each $\varepsilon_{pt}$ and $\Delta\varepsilon$, we utilized $100$ DNNs to train the EOSs of neutron stars, incorporating the phase transition, yielding $100$ sets of $\langle c_{s,i}^2\rangle$ at various energy density segments. Consequently, the Pearson correlation coefficients among different $\langle c_{s,i}^2\rangle$  from the $100$ DNNs can be computed. These correlations are depicted in Fig. \ref{fig9} for different $\varepsilon_{pt}$ and $\Delta\varepsilon$. The correlations of the speed of sound at energy density segments encompassing the phase transition with others are denoted as 'nan'. Generally, the correlations between different $\langle c_{s,i}^2\rangle$ are notably weak. They lack strong linear correlations, indicating the validity of our choice with the piecewise-polytropic representation. Moreover, it is observed that the speed of sound at the second segment after the phase transition exhibits the weakest correlation with other $\langle c_{s,i}^2\rangle$ for $\Delta=1.0\varepsilon$, whereas it is the one at the third segment after the phase transition for smaller $\Delta=0.5\varepsilon$.

\section{Summary}\label{sec4}

We utilized deep neural networks to delve into first-order phase transitions in neutron stars using the Maxwell construction method. The phase transition region was chosen within the energy density range of $2$-$4\varepsilon_0$, where $\varepsilon_0$ denotes the energy density at nuclear saturation point. The onset energy density and interval of phase transition were assumed as $\varepsilon_{pt}=2.0\varepsilon_0$, $2.5\varepsilon_0$, and $3.0\varepsilon_0$, with $\Delta\varepsilon=0.5\varepsilon_0$ and $1.0\varepsilon_0$, respectively. For each $\varepsilon_{pt}$ and $\Delta\varepsilon$, the EOSs of neutron stars including the phase transition were generated by DNNs comprising five layers. The input layer consisted of masses, radii, and their derivatives from observations of $17$ neutron stars. The output layers incorporated the average speed of sound, $\langle c^2_s\rangle$, which could represent the EOS in a piecewise-polytropic form. DNNs were trained using a dataset containing five million data points obtained by randomly {within the uniform distribution} selecting mass, radius, and their derivatives of neutron stars.

The EOS of neutron stars containing the phase transition was generally stiffer than that without phase transition before the appearance of quarks. Upon completion of the phase transition, the EOS of pure quark matter increased rapidly and eventually became stiffer than that of pure hadronic matter. These EOSs can result in heavier neutron stars, larger radii at phase transition segments, and greater dimensionless tidal deformabilities, thus meeting constraints imposed by observations including massive neutron stars, NICER, and gravitational waves. However, in the case that both the onset energy density and interval of phase transition are minimal, i.e., $\varepsilon_{pt}=2.0\varepsilon_0$ and $\Delta\varepsilon=0.5\varepsilon_0$, the EOSs with phase transitions tend to be intrinsically softer than that not considering phase transition. This leads to the derivation of neutron stars with lighter masses and smaller radii, in consistent with data from HESS J1731-347. Furthermore, the dimensionless tidal deformability is significantly reduced. 

In neutron stars devoid of phase transition, the trace anomalies will experience a monotonous decrease concomitant with an increase in density. In contrast, in areas undergoing phase transition, increments become apparent due to the pressures remaining unchanged during the first-order phase transition. In the present work, the real speed of sound, $c^2_s$, was strictly constrained to be less than $1$ for causality. With a phase transition, the average sound speed,  $\langle c^2_s\rangle$, was found to lie within the range of 0.2 and 0.7. At high-density regions, they approached the conformal limit, $1/3$. Furthermore, the average speeds of sound after the first-order phase transition were larger than those in pure hadronic matter. Our findings also indicated that the correlations of $\langle c^2_s\rangle$, derived from DNNs across varying energy density segments, were very weak.

Both with and without the phase transition, the EOSs from the DNNs  satisfied the constraints from present observables of neutron stars, such as mass, radius, and tidal deformability. Therefore, it poses a challenge to discern the specific onset energy and interval of phase transition from existing DNNs. Expectations incline toward the employment of high-precision probes, such as oscillations of neutron star, to investigate phase transitions in neutron stars within the framework of DNNs, providing more theoretical support for future gravitational wave facilities.

\section{Acknowledgments}
This work was supported in part by the National Natural Science Foundation of China	 No. 12175109, the Natural Science Foundation of Tianjin (Grant  No: 19JCYBJC30800), and the Natural Science Foundation of Guangdong Province (Grant  No: 2024A1515010911).  

\bibliographystyle{apsrev4-1}
\bibliography{refer}

\end{document}